%% file: main.tex
\documentclass[%
 reprint,
superscriptaddress,
bibnotes,
tightenlines,
notitlepage,
longbibliography,
twocolumn,
 amsmath,amssymb,
 aps,
]{revtex4-1}

\usepackage{graphicx}
\usepackage{dcolumn}
\usepackage{bm}
\usepackage[usenames,dvipsnames,svgnames,table]{xcolor}
\usepackage[colorlinks = true,
            linkcolor = Navy,
            urlcolor  = Navy,
            citecolor = Navy,
            anchorcolor = Navy]{hyperref}
\usepackage[caption=false]{subfig}

\usepackage{feynmf}

\usepackage{lineno}

\begin{document}

\input{side/title.tex}

\input{body/intro.tex}

\input{body/sim_det.tex}

\input{body/analysis.tex}

\input{body/sensitivity.tex}

\input{body/conclusion.tex}

\input{side/acknow.tex}

\bibliography{bibs/pheno.bib,bibs/software.bib,./bibs/exp.bib,./bibs/theory.bib,./bibs/upc.bib}

\end{document}

%% file: side/title.tex
\title{ Photon collider search strategy for sleptons and dark matter at the LHC }%

\author{Lydia Beresford}
 \email{lydia.beresford@physics.ox.ac.uk}
 \author{Jesse Liu}
 \email{jesse.liu@physics.ox.ac.uk}
\affiliation{%
 Department of Physics, University of Oxford, Oxford OX1 3RH, UK
}

\begin{abstract}
We propose a search strategy using the LHC as a photon collider to open sensitivity to scalar lepton (slepton $\tilde{\ell}$) production with masses around 15 to 60 GeV above that of neutralino dark matter $\tilde{\chi}^0_1$. This region is favored by relic abundance and muon $(g-2)_\mu$ arguments. However, conventional searches are hindered by the irreducible diboson background. We overcome this obstruction by measuring initial state kinematics and the missing momentum four-vector in proton-tagged ultraperipheral collisions  using forward detectors. We demonstrate sensitivity beyond LEP for slepton masses of up to 220~GeV for $ 15 \lesssim \Delta m(\tilde{\ell}, \tilde{\chi}^0_1) \lesssim 60$~GeV with 100~fb$^{-1}$ of 13~TeV proton collisions. We encourage the LHC collaborations to open this forward frontier for discovering new physics.
\end{abstract}

\maketitle

%% file: body/intro.tex
\section{Introduction}

Elucidating the elementary properties of dark matter (DM) is among the most urgent problems in fundamental physics. The lightest neutralino $\tilde{\chi}^0_1$ in supersymmetric (SUSY) extensions of the Standard Model (SM) is one of the most motivated DM candidates~\cite{Goldberg:1983nd,Ellis:1983ew,Bertone:2004pz}. A favored scenario involves scalar partners of the charged leptons (sleptons $\tilde{\ell}$) being one to tens of GeV above the $\tilde{\chi}^0_1$ mass. This enables interactions that reduce the $\tilde{\chi}^0_1$ cosmological relic abundance to match the observed value~\cite{Aghanim:2018eyx} via a mechanism called slepton coannihilation~\cite{Griest:1990kh,Edsjo:1997bg}. Furthermore, partners of the muon (smuon $\tilde{\mu}$) and neutralinos with masses near the weak scale are a leading explanation for $3-4\sigma$ deviations between measurements of the muon magnetic moment and SM prediction~\cite{Bennett:2006fi,Aoyama:2012wk,Hagiwara:2011af,Ajaib:2015yma}. 

Remarkably, Large Hadron Collider (LHC) searches for these key targets have no sensitivity when mass differences are $15 \lesssim \Delta m(\tilde{\ell}, \tilde{\chi}^0_1) \lesssim 60$~GeV~\cite{Aad:2014vma,Aaboud:2018jiw,Sirunyan:2018nwe,Aaboud:2017leg}. Here, Large Electron Positron (LEP) collider limits remain the most stringent, excluding $m(\tilde{\ell}) \lesssim 97$~GeV~\cite{LEPlimits_slepton,Heister:2001nk,Achard:2003ge}. Sensitivity is hindered by an obstruction generic to all LHC search strategies for invisible DM states and their mediators~\cite{Alwall:2008ag,Alwall:2008va,Alves:2011wf,Alves:2011wf,Essig:2011qg,Buchmueller:2014yoa,Abdallah:2015ter,Abercrombie:2015wmb,Aaboud:2017phn,ATLAS:2015nsi,Aaboud:2017dor,Aaboud:2017yqz,Dutta:2017nqv}: the kinematics of colliding quarks and gluons are immeasurable. Without this initial state information, the missing momentum four-vector $p_\text{miss}$ left by DM can only be determined in the plane transverse to the beam ($\mathbf{p}_\text{T}^\text{miss}$). This precludes direct DM mass reconstruction that would otherwise provide effective discrimination against neutrino $\nu$ backgrounds.

\begin{figure}[tbp]
        \includegraphics[width=0.45\columnwidth]{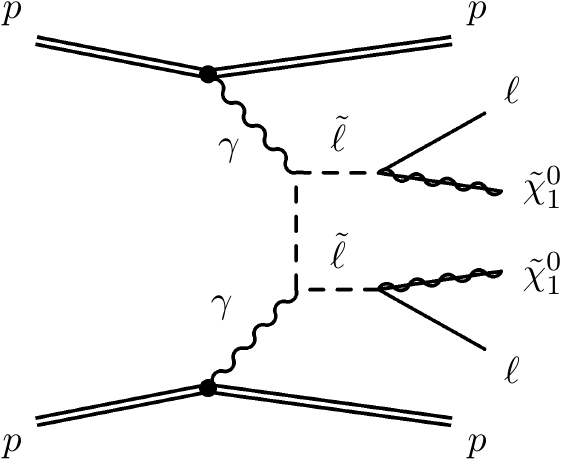}
        \includegraphics[width=0.45\columnwidth]{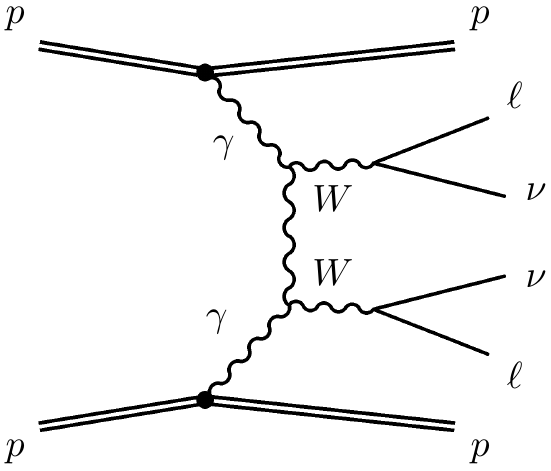}
    \caption{\label{fig:aaFeyn}Exclusive pair production of (left) scalar leptons `sleptons' $\tilde{\ell}$ decaying to dark matter $\tilde{\chi}^0_1$ and (right) SM diboson $WW$ background using the LHC as a photon collider. 
    }
\end{figure}

This Letter proposes a search strategy to resolve these longstanding problems by using the LHC as a photon collider~\cite{Piotrzkowski:2000rx}. In a beam crossing, protons can undergo an ultraperipheral collision (UPC), where photons from the electromagnetic fields interact to produce sleptons exclusively $pp \to p(\gamma\gamma \to \tilde{\ell}\tilde{\ell}) p$. The sleptons decay as $\tilde{\ell} \to \ell \tilde{\chi}^0_1$, resulting in the very clean final state $p (2\ell + p_\text{miss}) p$ of our search: two intact protons, two leptons $\ell$, and missing momentum (Fig.~\ref{fig:aaFeyn}). As the beam energy is known, measuring the outgoing proton kinematics determines the colliding photon momenta and thus $p_\text{miss}$. This experimental possibility is opened by the ATLAS Forward Proton (AFP)~\cite{Adamczyk:2017378} and CMS--TOTEM Precision Proton Spectrometer (CT-PPS)~\cite{Albrow:1005180,Albrow:1753795} forward detectors, which recorded first data in 2017 and 2016 respectively. CMS--TOTEM moreover observed double lepton production in high-luminosity proton-tagged events~\cite{Cms:2018het}, demonstrating  initial state reconstruction is feasible. 

Photon collisions at the LHC reach sufficient rates to probe rare processes such as SM light-by-light scattering~\cite{dEnterria:2013zqi,Aaboud:2017bwk}, anomalous gauge couplings~\cite{Chapon:2009hh,Fichet:2013gsa}, and axion-like particles~\cite{Knapen:2016moh,Baldenegro:2018hng}. Nonetheless, it is widely considered that photon fusion production of sleptons is not competitive as a discovery window compared to electroweak production~\cite{Aad:2014vma,Aaboud:2018jiw,Sirunyan:2018nwe,Aaboud:2017leg}; existing photon collider studies therefore focus on slepton mass measurement for specific benchmark points~\cite{Ohnemus:1993qw,Schul:2008sr,Albrow:2008pn,deFavereaudeJeneret:2009db,HarlandLang:2011ih}. Our proposal argues the contrary that photon collisions play an essential role in SUSY and DM searches. We emulate AFP/CT-PPS proton tagging, which enables powerful background suppression. We demonstrate a strategy that surpasses LEP sensitivity in the favored $15 \lesssim \Delta m(\tilde{\ell}, \tilde{\chi}^0_1) \lesssim 60$~GeV corridor, underscoring the importance of initial state kinematics and $p_\text{miss}$ for the LHC discovery program.

%% file: body/sim_det.tex
\section{\label{sec:sim_det} Photon collider simulation}

Electromagnetic fields surrounding ultrarelativistic protons can be modeled as a beam of nearly on-shell photons, which is known as the equivalent photon approximation~\cite{Budnev:1974de}. We consider pair production of electrically charged particles $X$ via photon fusion $\gamma\gamma \to XX$. Analytic expressions of their QED cross-sections $\sigma_{\gamma\gamma \to XX}$ may be found in Refs.~\cite{Brodsky:1971ud,PhysRevD.23.1933,Ohnemus:1993qw,HarlandLang:2011ih}. The LHC cross-section is then the convolution of $\sigma_{\gamma\gamma \to XX}$ with the effective photon luminosity $L_{\gamma\gamma}^{(pp)}$ from the protons   
\begin{equation}
    \sigma_{pp \to p (\gamma\gamma \to XX) p} = \int \sigma_{\gamma\gamma \to XX} (m_{\gamma\gamma}) \frac{\mathrm{d} L_{\gamma\gamma}^{(pp)}}{\mathrm{d} m_{\gamma\gamma}}\mathrm{d} m_{\gamma\gamma},
    \label{eq:xsec_pp_photon}
\end{equation}
where $m_{\gamma\gamma}$ is the invariant mass of the two-photon system. We use \textsc{MadGraph}~v2.6.1~\cite{Alwall:2011uj,Alwall:2014hca} to numerically evaluate Eq.~\eqref{eq:xsec_pp_photon} and perform Monte Carlo simulation for signal and background processes. We study the resulting events using the \textsc{pylhe} package~\cite{lukas_2018}, and parameterize detector effects as follows.

The forward detectors identify both the intact outgoing protons at $z\simeq \pm 220$~m downstream from the collision point and measure their energies $E_\text{forward}$. Protons are steered outside the beam profile by the LHC dipole magnets due to the fractional energy loss $\xi = 1- E_\text{forward}/E_\text{beam}$ relative to the beam energy $E_\text{beam}$. The AFP/CT-PPS proton acceptance is close to 100\% for $0.015 < \xi < 0.15$~\cite{Adamczyk:2017378,Albrow:1005180,Albrow:1753795}. This translates to emitted photon energies of $100 \lesssim E_\gamma \lesssim 1000$ GeV for $\sqrt{s} = 13$~TeV $pp$ collisions. The survival probability of a proton remaining intact following photon emission is reported to be around 90\% in phenomenology studies~\cite{Khoze:2001xm}, which we treat as an efficiency. We parameterize the proton acceptance as $\{0, 0.5, 0.7, 0.9\}$ for $E_\gamma \in \{[0, 100], [100, 120], [120, 150], [150, 400]\}$~GeV respectively, and 0.8 otherwise. We conservatively smear the photon four-vector $p_\gamma^\text{smeared} = p_\gamma^\text{generated} G_\gamma(1, \sigma_\gamma)$ using a Gaussian $G_\gamma$ with width $\sigma_\gamma = 5\%$, based on the AFP resolution of 5~GeV at $\xi \simeq 0.015$~\cite{Adamczyk:2017378}. 

The central detectors reconstruct isolated leptons (electrons $e$ and muons $\mu$ throughout). To emulate detector resolution, we smear the lepton momenta $p_\ell$ using a Gaussian $G_\ell$ with width $\sigma_\ell = 5\%$. We parameterize $p_\text{T}$-dependent reconstruction efficiencies in accord with ATLAS~\cite{Aaboud:2017leg}, which account for all lepton quality conditions. This requires that leptons satisfy transverse momentum $p_\text{T}^{e(\mu)} > 4.5(4)$~GeV and pseudorapidity $|\eta_\ell| < 2.5$.

To simulate the simplified model signal $\gamma\gamma \to \tilde{\ell}\tilde{\ell}$, we employ the model specified by the \texttt{SLHA} parameter file from the auxiliary material of Ref.~\cite{Aaboud:2017leg}. This allows comparisons with existing LHC constraints. Only sleptons $\tilde{\ell}$ and the stable neutralino $\tilde{\chi}^0_1$ are kinematically accessible, whose masses are free parameters. A fourfold mass degeneracy is assumed such that scalar partners of the left-handed and right-handed electrons and muons (selectrons $\tilde{e}$ and smuons $\tilde{\mu}$) satisfy $m(\tilde{\ell}_{L, R}) = m(\tilde{e}_L) = m(\tilde{e}_R) = m(\tilde{\mu}_L) = m(\tilde{\mu}_R)$. The sleptons decay $\tilde{\ell} \to \ell\tilde{\chi}^0_1$ with 100\% branching ratio and are handled by \textsc{MadGraph}. All other SUSY states are kinematically inaccessible with masses well above 10~TeV. We sample $m(\tilde{\ell})$ in 25 GeV steps, and $\Delta m(\tilde{\ell}, \tilde{\chi}^0_1)$ in steps of no more than 20~GeV. We simulate 50k events per mass point and normalize to cross-sections calculated in \textsc{MadGraph}, which are consistent with those obtained in Refs.~\cite{deFavereaudeJeneret:2009db,HarlandLang:2011ih}. For $m(\tilde{\ell}) = 100$~GeV, the cross-section is 2.5~fb and falls to 0.25~fb for $m(\tilde{\ell}) = 200$~GeV. Only the first two generations $\tilde{\ell} \in [\tilde{e}, \tilde{\mu}]$ are considered; study of scalar partners of tau leptons (staus $\tilde{\tau}$) are deferred to future work.

%% file: body/analysis.tex
\section{\label{sec:analysis} Search strategy}

\begin{figure*}
        \includegraphics[width=0.32\textwidth]{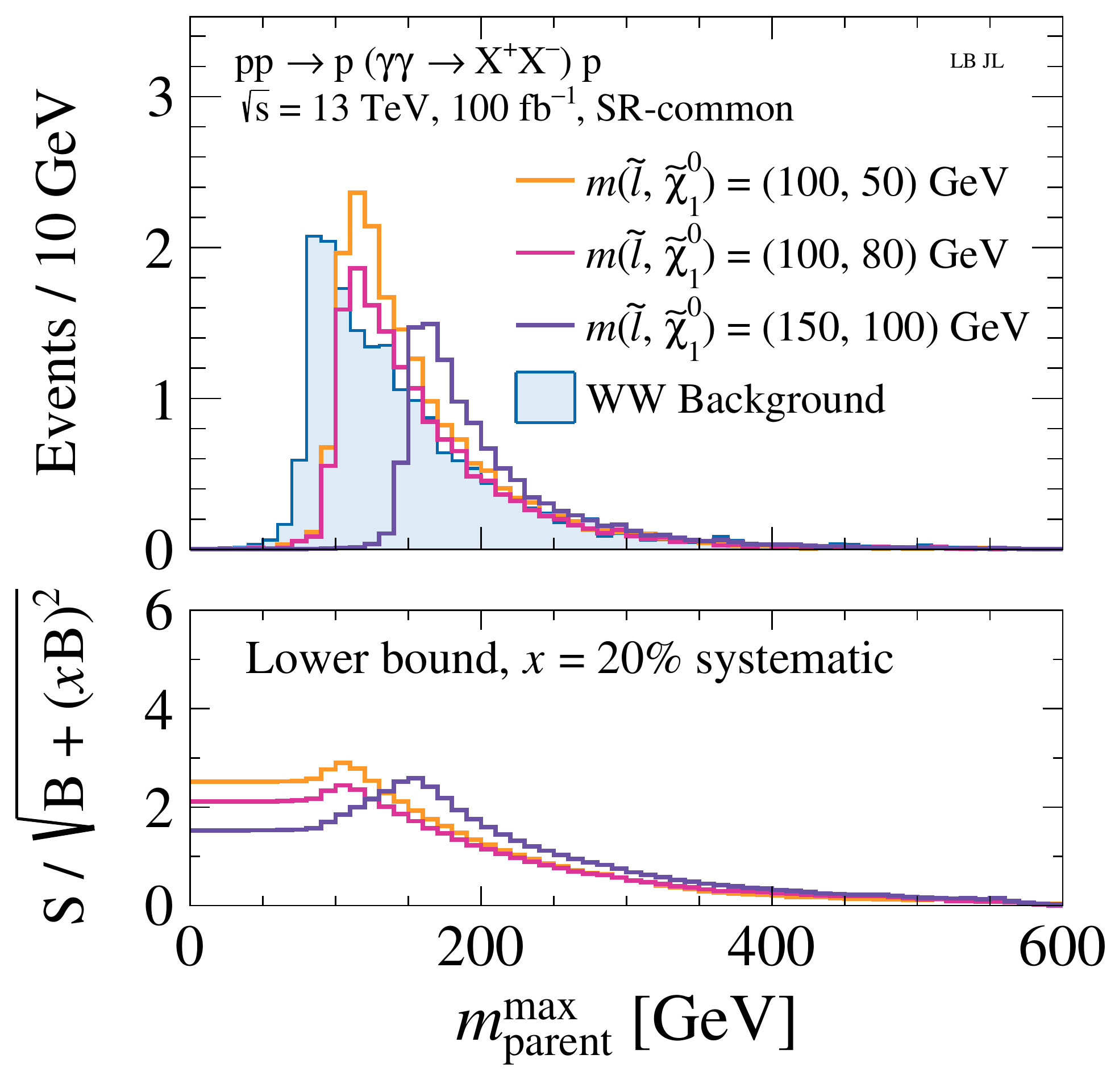}
        \includegraphics[width=0.32\textwidth]{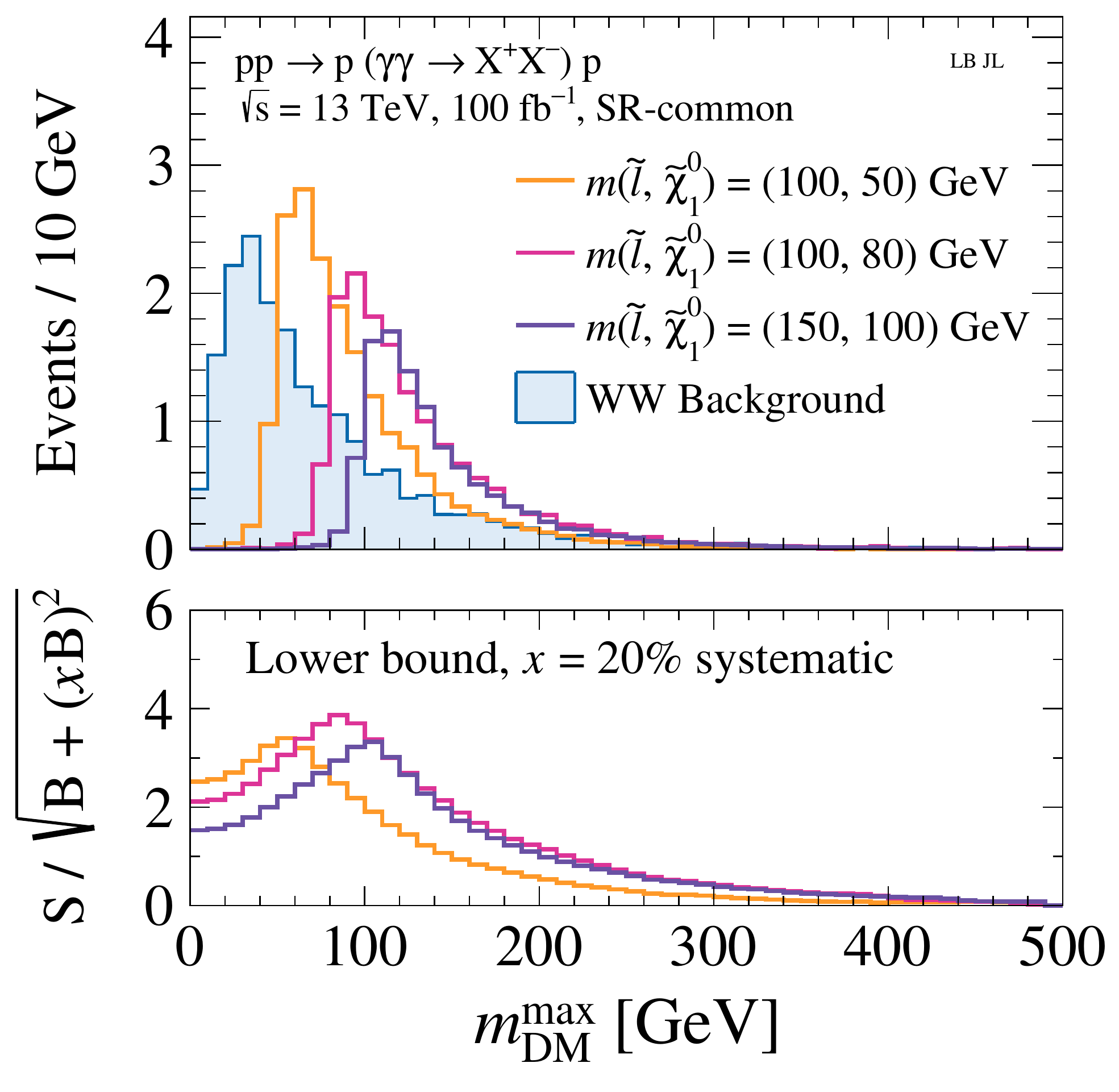}
        \includegraphics[width=0.32\textwidth]{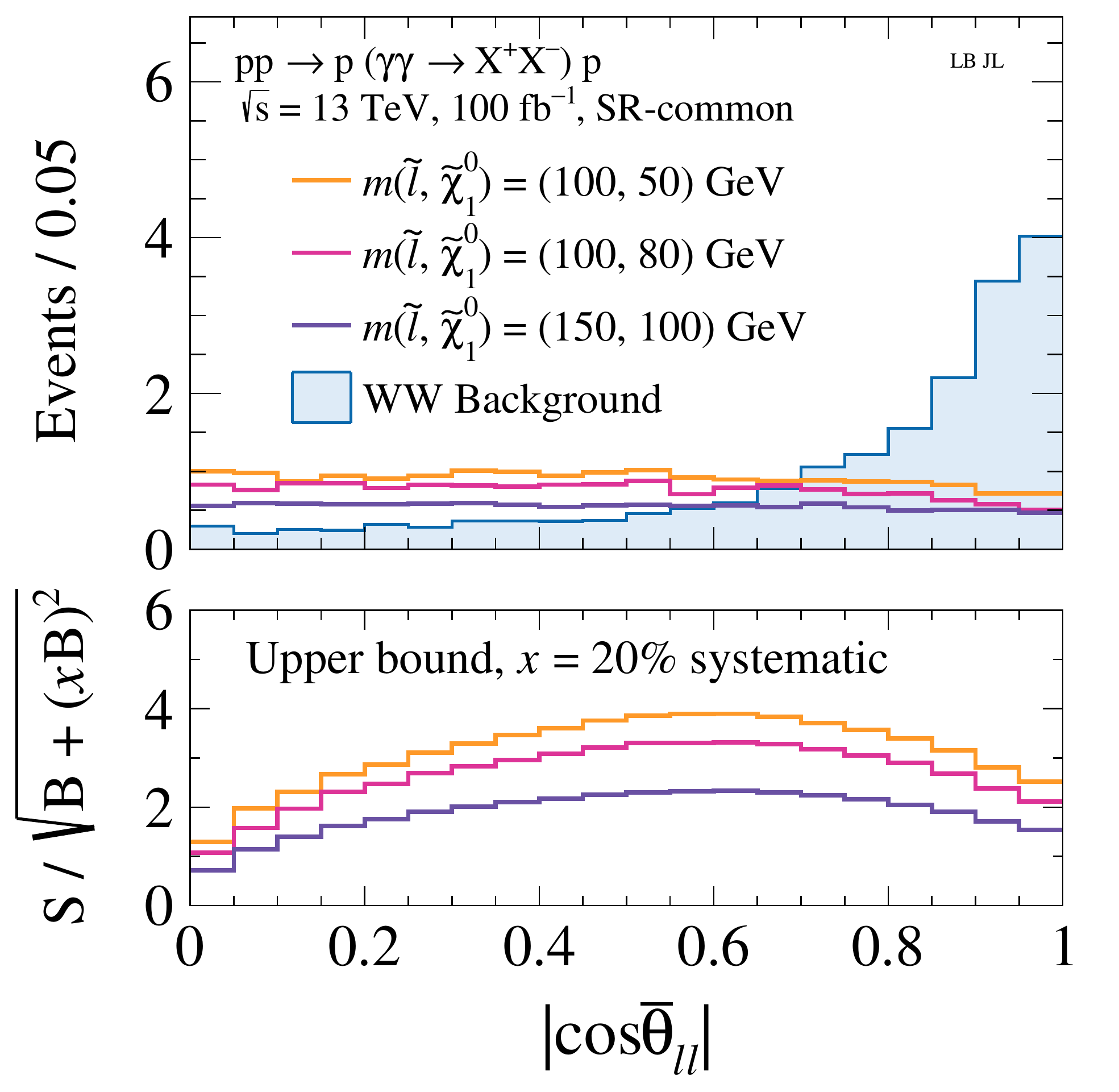}
    \caption{\label{fig:mass_vars} Kinematic distributions of search discriminants reconstructing the mass and spin of benchmark slepton signals (lines) and $WW$ background (filled), normalized to $100$~fb$^{-1}$. Double proton tag, lepton efficiencies and detector smearing are applied, but no lepton trigger emulation is imposed. The event selection applied, denoted SR-common, requires $m_\text{DM}^\text{max} > 0$~GeV, $|\eta_\ell| < 2.5$, same flavour leptons, and $m_\text{T2} > 2$~GeV. Masses of the signals are displayed in the legend. The lower panel estimates the statistical significance after integrating the signal $S$ and background $B$ counts with the indicated bound on the variable. }
\end{figure*}

Our search strategy focuses on extracting the signal from the dominant irreducible $\gamma\gamma \to WW \to \ell\nu\ell\nu$ background. The $WW$ cross-section times dileptonic branching fraction $\mathcal{B}$ is $\sigma_{\gamma\gamma \to WW} \times \mathcal{B} \simeq 5$~fb, which is comparable in size to the slepton signals. We generate 50k events of this process using \textsc{MadGraph}, which also handles the decays to preserve spin correlations of the leptons. We use dilepton triggers for event selection, which we emulate using a $p_\text{T}^\ell > 15$~GeV condition. Requiring same flavour leptons ($ee$ or $\mu\mu$) halves the $WW$ background while preserving signal. We then reconstruct three defining features that characterize the signals and background to optimize search sensitivity: mediator mass ($W$ or $\tilde{\ell}$), invisible mass ($\nu$ or $\tilde{\chi}^0_1)$, and mediator spin. 

At the LHC, proton-tagging enables unambiguous bounds on both the parent mediator and DM masses. The mass of the $\tilde{\ell}$ mediators is bound by the invariant mass of the initial state two-photon system $
    m_{\gamma\gamma}^2 = (p_{\gamma_1} + p_{\gamma_2})^2 \geq (2 m_{\tilde{\ell}})^2$.
Meanwhile, the invariant mass of the invisible system $W_\text{miss}$ bounds the DM masses
$W_\text{miss}^2 = p_\text{miss}^2 \geq (2 m_{\tilde{\chi}^0_1})^2$. Here, $p_\text{miss} = \sum_i p_i - \sum_f p_f$ is the vectorial sum of the momenta of the visible final states $p_f$ subtracted from the initial states $p_i$. In this search, we have $ \sum_i p_i = p_{\gamma_1} + p_{\gamma_2}$ and $ \sum_f p_f = p_{\ell_1} + p_{\ell_2}$. We find the ratio $m_{\gamma\gamma} / W_\text{miss}$ to be useful for $\Delta m(\tilde{\ell}, \tilde{\chi}^0_1) \lesssim 30$~GeV signals. 

To improve mass reconstruction of the parent mediator and DM states, one can impose hypotheses about the decay topology. Assuming the symmetric pair of semi-invisible decays $\tilde{\ell}\tilde{\ell} \to \ell\tilde{\chi}^0_1 \ell\tilde{\chi}^0_1$, with photon and lepton momenta measured, results in the HKSS variables~\cite{HarlandLang:2011ih}. These also provide mass bounds on the parent mediator and invisible system (see Ref.~\cite{HarlandLang:2011ih} for definition)
\begin{linenomath}
\begin{align}
m_\text{parent}^\text{max}  \geq m(\tilde{\ell}),\quad
m_\text{DM}^\text{max}      \geq m(\tilde{\chi}^0_1).
\end{align}
\end{linenomath}
Importantly, these variables have more steeply falling tails than $m_{\gamma\gamma}$ and $W_\text{miss}$ respectively, and therefore provide better signal separation from the $WW$ background. 

To exploit the mediator spin for sensitivity, we use the Barr--Melia variable~\cite{Barr:2005dz,Melia:2011cu}, defined by 
\begin{linenomath}
\begin{equation}
\cos \bar{\theta}_{\ell\ell} = \tanh\left[\tfrac{1}{2}(\bar{\eta}_{\ell_1} - \bar{\eta}_{\ell_2})\right],
\end{equation}
\end{linenomath}
where the pseudorapidities $\bar{\eta}$ are evaluated in the dilepton centre-of-mass frame (denoted by overlines). Leptons from spin 0 $\tilde{\ell}$ mediators decay more centrally than those from spin 1 $W$ bosons, offering discrimination power. 

Figure~\ref{fig:mass_vars} displays distributions of benchmark signals and the $WW$ background for these mass and spin sensitive variables, normalized to 100~fb$^{-1}$. From this, we impose $|\cos\bar{\theta}_{\ell\ell}| < 0.65$ and construct three signal region (SR) categories targeting small `compressed', medium `corridor', and large mass differences $\Delta m(\tilde{\ell}, \tilde{\chi}^0_1)$:
\begin{itemize}
\item SR-compressed: $m_\text{parent}^\text{max} > 80$~GeV, $m_\text{DM}^\text{max} > 0$~GeV, $m_{\gamma\gamma} / W_\text{miss} < 1.4 $;
\item SR-corridor: $m_\text{parent}^\text{max} > 120$~GeV, $m_\text{DM}^\text{max} > 80$~GeV;
\item SR-large: $m_\text{parent}^\text{max} > 130$~GeV, $m_\text{DM}^\text{max} > 20$~GeV. 
\end{itemize}
An improved search strategy would involve a shape analysis across $m_\text{parent}^\text{max}$ vs $m_\text{DM}^\text{max}$ akin to a bumphunt~\cite{ATLAS:2015nsi} in two dimensions, but is beyond the scope of this work. 

Other potential irreducible processes include $\tau\tau \to \ell \nu\nu \ell\nu\nu$, which has a large rate $\sigma \times \mathcal{B} \simeq 74 \times 0.35^2 \simeq 9.1~$pb. We reject this process by reconstructing the $\tau$ mass endpoint using the stransverse mass $m_\text{T2} > 2$~GeV (see Refs.~\cite{Lester:1999tx,Barr:2003rg,Lester:2014yga} for definition). This variable uses the lepton momenta and missing transverse momentum defined by $\mathbf{p}_\text{T}^\text{miss} \equiv - \mathbf{p}_\text{T}^{\ell_1} - \mathbf{p}_\text{T}^{\ell_2} $. We validate mitigation of this background by generating an event sample in \textsc{MadGraph} using the \texttt{sm-lepton\_masses} model to decay the taus. Top quark pairs $\gamma\gamma \to t\bar{t} \to b\ell\nu b\ell\nu$ contribute a small rate $\sigma \times \mathcal{B} \simeq 0.33 \times 0.21^2 \simeq 0.015$~fb and we assume a jet veto renders this background negligible. 

Turning to reducible backgrounds induced by detector misreconstruction, these typically require data-driven techniques by the experimental collaborations to estimate reliably. We briefly discuss possible mitigation strategies. First, nonresonant production of lepton pairs $\gamma\gamma \to \ell\ell$, where $\ell \in [e, \mu]$, has a large cross-section of 140~pb per flavour. Missing momentum results solely from detector resolution and this background is also rendered negligible by the $m_\text{T2}$ requirement. This also suppresses resonant dilepton processes from decays of diquark bound states, such as $J/\psi$ and $\Upsilon$ resonances. Next, leptons from fake and nonprompt sources, such as semileptonic decays of $B$-hadrons, typically become significant at low lepton $p_\text{T}$~\cite{Aaboud:2017leg}. We expect these to be well controlled by standard lepton quality requirements in the extremely clean events. Finally, protons from pileup collisions can fake intact UPC protons when occurring in the same event as an exclusive or nonexclusive process that gives two leptons and $p_\text{miss}$. A veto in the Zero Degree Calorimeter~\cite{ATLAS:2007aa} will suppress nonexclusive processes. Timing with 10~ps resolution can associate protons in the forward detectors to the lepton vertices~\cite{Royon2014}.

%% file: body/sensitivity.tex
\section{\label{sec:sensitivity} Sensitivity and discussion}

We now evaluate the sensitivity of our search strategy for the slepton--DM simplified model. We assume two benchmark luminosities $\mathcal{L} = 100~(300)$~fb$^{-1}$, which correspond to the cumulative dataset for LHC Run~2 (3). We use the asymptotic Poisson significance with uncertain background $Z_A(S, B, \sigma_B)$~\cite{Cowan:poissonunc,Cowan:2010js}. This takes as input the signal $S$, background $B$ counts, and we ascribe a background systematic uncertainty of $\sigma_B = 0.2 B$. For 100~fb$^{-1}$, SR-compressed has $B=0.55$ and the highest $S=6.7$ is for the $m(\tilde{\ell}, \tilde{\chi}^0_1) = (100, 80)$~GeV signal. This corresponds to a signal efficiency of 2.6\%, and a significance of $3.9\sigma$, rising to $6.8\sigma$ for 300~fb$^{-1}$. Meanwhile, SR-corridor targets slightly larger $\Delta m(\tilde{\ell}, \tilde{\chi}^0_1)$, where $B = 1.1$ and the highest $S = 8.1$ corresponds to the $m(\tilde{\ell}, \tilde{\chi}^0_1) = (125,80)$~GeV signal, translating to $4.0\sigma$ significance, rising to $6.9\sigma$ for 300~fb$^{-1}$. SR-large probes larger $\Delta m(\tilde{\ell}, \tilde{\chi}^0_1)$, with $B=1.6$ at 100~fb$^{-1}$ and the highest $S=8.5$ is for the $m(\tilde{\ell}, \tilde{\chi}^0_1) = (125, 40)$~GeV signal.

\begin{figure}
    \centering
    \includegraphics[width=\columnwidth]{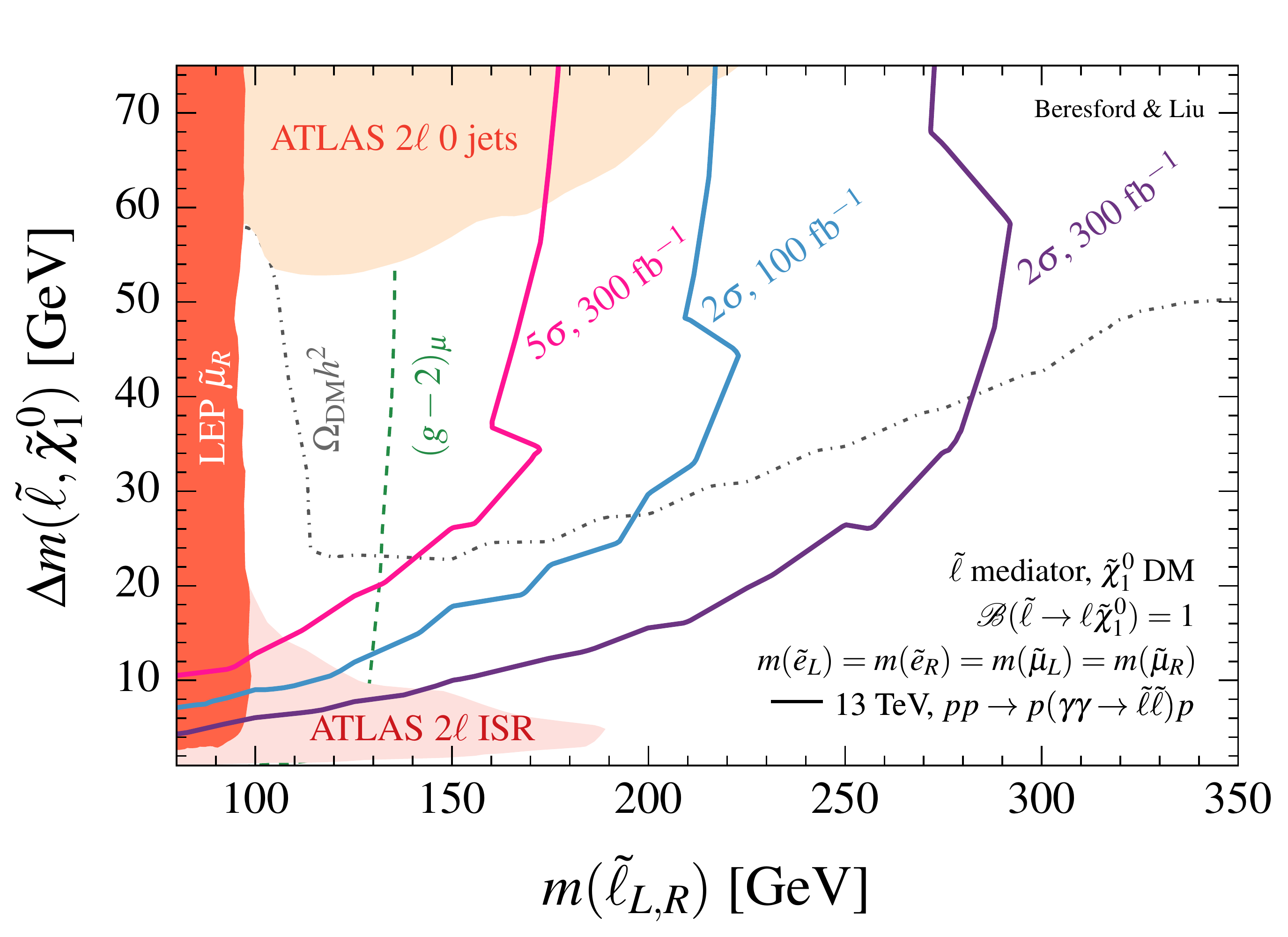}
    \caption{Projected photon collider sensitivity of $\gamma\gamma \to \tilde{\ell}\tilde{\ell}$ using 13 TeV proton-tagged LHC collisions. Solid lines (this work) show the $2\sigma$ sensitivity contours for integrated luminosities of 100~fb$^{-1}$ (blue) and 300~fb$^{-1}$ (purple), along with $5\sigma$ at 300~fb$^{-1}$ (pink). A simplified model of slepton mediators $\tilde{\ell}$ with a fourfold mass degeneracy decaying to neutralino DM $\tilde{\chi}^0_1$ is considered. 
    Filled regions denote constraints from ATLAS $2\ell$ 0 jets~\cite{Aad:2014vma,Aaboud:2018jiw} (yellow), $2\ell$ ISR searches~\cite{Aaboud:2017leg} (pink), and LEP for partners of the right-handed muons $\tilde{\mu}_R$~\cite{LEPlimits_slepton,Heister:2001nk,Achard:2003ge} (orange).
    Dashed lines indicate parameter space favored by relic abundance $\Omega_\text{DM} h^2$~\cite{Aghanim:2018eyx} (gray) and muon $(g-2)_\mu$~\cite{Aoyama:2012wk} (green) measurements, computed using \textsc{micrOMEGAs}~\cite{Belanger:2014vza}. }
    \label{fig:slepton_summary}
\end{figure}

Figure~\ref{fig:slepton_summary} shows the $2\sigma$ `sensitivity' contours of our search strategy (solid lines) in the $\Delta m(\tilde{\ell}, \tilde{\chi}^0_1)$ vs $m(\tilde{\ell})$ plane, with $5\sigma$ `discovery' contours displayed for 300~fb$^{-1}$. For each signal point, we use the highest significance out of the three SRs. Our strategy unambiguously surpasses existing collider sensitivity (filled regions) in the $15\lesssim \Delta m(\tilde{\ell}, \tilde{\chi}^0_1) \lesssim 60$~GeV corridor. For $\Delta m(\tilde{\ell}, \tilde{\chi}^0_1) \sim 40$~GeV, $2\sigma$ sensitivity reaches $m(\tilde{\ell}) \sim 220~(280)$~GeV for 100 (300)~fb$^{-1}$, while $5\sigma$ sensitivity extends up to $m(\tilde{\ell}) \sim 160$~GeV using 300~fb$^{-1}$. 

The mass reach depends on several factors. As $m(\tilde{\ell})$ increases, the $\gamma\gamma \to \tilde{\ell}\tilde{\ell}$ cross-section decreases and the search becomes statistically limited. However, signals with larger $m(\tilde{\ell})$ are easier to distinguish from the $WW$ background as the signal becomes better separated from the $W$ boson mass; higher DM masses are similarly easier to separate. For $m(\tilde{\ell}) \lesssim 130$~GeV, sensitivity is limited by the forward detector acceptance, which drops rapidly for proton energy losses of $E_\gamma \lesssim 100$~GeV.  

The canonical LHC search for sleptons employs the `$2\ell$ 0 jets' signature, where the ATLAS 8 TeV, 20.3~fb$^{-1}$ analysis gives the most stringent limit for $m(\tilde{\ell}) \lesssim 250$~GeV~\cite{Aad:2014vma}. Notably, the 13 TeV, 36.1~fb$^{-1}$ counterpart~\cite{Aaboud:2018jiw} did not surpass the 8~TeV analysis sensitivity for $\Delta m(\tilde{\ell}, \tilde{\chi}^0_1) \lesssim 60$~GeV, despite higher centre-of-mass energy and luminosity, with similar results from CMS~\cite{Sirunyan:2018nwe}.

Our strategy has limited sensitivity to the very compressed region $\Delta m(\tilde{\ell}, \tilde{\chi}^0_1) \lesssim 10$~GeV due to the trigger emulation $p_\text{T}^{\ell} > 15$~GeV. Recent work proposed strategies using initial state radiation (ISR) and low momentum leptons to probe this challenging region~\cite{Han:2014aea,Barr:2015eva}, which is successfully adopted by the ATLAS $2\ell$ ISR search~\cite{Aaboud:2017leg}. Our strategy can potentially provide a complementary probe of this region, free from hadronic backgrounds. This is only possible if lepton trigger thresholds are lowered by using forward detector triggering, motivating their development for LHC Run~3.

A striking feature of Fig.~\ref{fig:slepton_summary} is that our proposal decisively probes regions favored by DM and muon $(g-2)_\mu$ phenomenology. We evaluate these noncollider observables using \textsc{micrOMEGAs}~v4.2.1~\cite{Belanger:2014vza}. The gray dashed contour indicates where the $\tilde{\chi}^0_1$ relic abundance matches the Planck measurement $\Omega_{\tilde{\chi}^0_1} h^2 = \Omega_\text{DM}^\text{Planck} h^2 = 0.12$~\cite{Aghanim:2018eyx}. Depletion of $\Omega_{\tilde{\chi}^0_1} h^2$ occurs via coannihilation processes such as $\tilde{\ell}\tilde{\chi}^0_1 \to \ell \gamma$, whose rate grows exponentially $\sim e^{-\Delta m(\tilde{\ell}, \tilde{\chi}^0_1)/m(\tilde{\ell})}$ with smaller mass differences~\cite{Griest:1990kh,Edsjo:1997bg}. At low $m(\tilde{\ell})$, the self-annihilation via the $Z$ boson `funnel' becomes competitive, allowing larger mass splittings to satisfy $\Omega_\text{DM}^\text{Planck} h^2$. Loop corrections from $\tilde{\ell}$ and $\tilde{\chi}^0_1$ states contribute to the muon anomalous magnetic moment $a_\mu = \frac{1}{2}(g-2)_\mu$. The green dashed line indicates modifications consistent with the measured discrepancy $\Delta a_\mu = a_\mu^\text{measured} - a_\mu^\text{predicted} \simeq 2.5\times 10^{-9}$~\cite{Aoyama:2012wk}. While we consider these features in a simplified model, the phenomenology is qualitatively consistent with those in global fits of more complete 11-parameter models~\cite{Bagnaschi:2017tru}.

If the fourfold mass degeneracy scheme is relaxed, the LHC blind corridor widens to $10\lesssim \Delta m(\tilde{\mu}_R, \tilde{\chi}^0_1) \lesssim 90$~GeV~\cite{Aad:2014vma,Aaboud:2018jiw,Sirunyan:2018nwe,Aaboud:2017leg}, where our strategy will play an important role. In conventional electroweak production, the right-handed states $\tilde{\ell}_R$ have order 3 times smaller cross-sections than the left-handed $\tilde{\ell}_L$ counterparts~\cite{Fuks:2013lya}. By contrast, the photon collider strategy has the advantage of equal QED cross-sections for $\tilde{\ell}_L$ and $\tilde{\ell}_R$ states.

%% file: body/conclusion.tex
This proposal is widely extendable to other search channels and electrically charged targets. So-called R-parity violating scenarios where the $\tilde{\chi}^0_1$ decays to higher multiplicity final states can profit from clean events. Charged fermions (charginos) face similar difficulties discriminating against $WW$ backgrounds and may benefit in combination with a hadronic channel. Scalar quarks, charged Higgs bosons, spin 1 mediators, disappearing track signatures are also motivated scenarios. 

In summary, we proposed a search strategy using the LHC as a photon collider to open sensitivity beyond LEP in the challenging corridor $15 \lesssim \Delta m(\tilde{\ell}, \tilde{\chi}^0_1) \lesssim 60$~GeV favored by DM and $(g-2)_\mu$ phenomenology. Proton tagging enables the initial state and missing momentum four-vector $p_\text{miss}$ to be reconstructed, offering striking background discrimination inaccessible to current LHC searches. We encourage experimental collaborations to include this forward physics frontier in flagship hadron collider searches for DM and their charged mediators.

%% file: side/acknow.tex

\emph{Acknowledgements}---We thank the hospitality of the DM@LHC Workshop at Heidelberg University, where discussions for this work began. We are grateful to Alan Barr for helpful conversations and feedback on the manuscript. LB is supported by St John's College, Oxford. JL is supported by STFC. 